\documentclass[a4paper]{cas-sc}
\immediate\write18{%
	makeindex -s nomencl.ist -o \jobname.nls -t \jobname.nlg \jobname.nlo%
}
 
\usepackage[numbers,sort]{natbib}
\usepackage{floatrow}
\usepackage{placeins}
\usepackage{xcolor}

\usepackage{amsmath}
\usepackage{lipsum}
\usepackage{resizegather}
\usepackage{textcomp}
\usepackage{multicol}
\usepackage{mathtools}
\usepackage{multirow}
\usepackage{enumitem}
\usepackage{tabularx,booktabs}
\usepackage{tabularx}
\usepackage{cuted}
\usepackage{tcolorbox}
\usepackage{csquotes}
\usepackage{subcaption}
\usepackage{nomencl}
\usepackage{etoolbox}
\usepackage{multibib}
 \usepackage{graphicx}
 \usepackage[justification=centering]{caption}
\usepackage{subcaption}
\usepackage{framed} 
\makenomenclature
\renewcommand*\nompreamble{\begin{multicols}{2}}
\renewcommand*\nompostamble{\end{multicols}}
\usepackage{etoolbox}
\renewcommand\nomgroup[1]{%
	\item[\bfseries
	\ifstrequal{#1}{I}{Indices and Sets }{%
		\ifstrequal{#1}{P}{Parameters and Constants  }{%
			\ifstrequal{#1}{F}{Functions and Variables }{%
			\ifstrequal{#1}{S}{ Symbols}}}}%
	]}
\graphicspath{{../PEMPicx/}}
\DeclareGraphicsExtensions{.pdf,.jpeg,.png,.eps,.tiff}
\begin{document}
\let\WriteBookmarks\relax
\shorttitle{}
\shortauthors{}

\title [mode = title]{An Adaptive Demand Response Framework using Price Elasticity Model in Distribution Networks: A Case Study 
}                      

%
%
%

\author{\textcolor{black}{Vipin Chandra Pandey}}
\author{ \textcolor{black}{Nikhil Gupta}}
\author{\textcolor{black}{K.~R.~Niazi}}
\author{\textcolor{black}{Anil Swarnkar}}
\author{\textcolor{black}{Rayees Ahmad Thokar}}
%
%
\address{Department of Electrical Engineering, 
Malaviya National Institute of Technology, Jaipur, India 
}
%
%
%
%
%
%
%
%

\begin{abstract}
Price elasticity model (PEM) is an appealing and modest model for assessing the potential of flexible demand in DR. It measures the customer’s demand sensitivity through elasticity in relation to price variation. However, application of PEM in DR is partially apprehensible on attributing the adaptability and adjustability with intertemporal constraints in DR. Thus, this article presents an adaptive economic DR framework with attributes of DR via a dynamic elasticity approach to model customer’s sensitivity. This dynamic elasticity is modeled through the deterministic and stochastic approaches. Both approaches envision the notion of load recovery for shiftable/flexible loads to make the proposed DR framework adaptive and adjustable relative to price variation. In stochastic approach, a geometric Brownian motion is employed to emulate load recovery with inclusion of intertemporal constraint of load flexibility. The proposed mathematical model shows what should be the customers’ elasticity value to achieve the factual DR. The case study is carried out on standard IEEE 33 distribution system bus load data to assess technical and socio-economic impact of DR on customers and is also compared with the exiting model.    
\end{abstract}



\maketitle

\section{Introduction}

The fusion of new resources and technological breakthrough has transformed the distribution network operation and control. Many thanks to smart grids, which have infused distribution network with new resources that exhibit dual nature (i.e., source and sink). In this, energy storage, electric vehicle to grid or vice-versa and prosumers (consumers which produces and consumes power themselves) are newly evolved state-of-the-art technologies in the distribution network~\cite{Chao2011b}. These transformations have brought advantages and also challenges in the system operation and control. In the meanwhile, demand response (DR) also has made its presence felt in the system. It is a result of persistent events such as sudden peak rise, increased penetration of renewable generation, wholesale price fluctuation, uncharacteristic dynamic behaviour of loads, reliability and security issues prevailing in the system~\cite{Schweppe1988,OfEnergy2006}. In the past few decades, there have been a lot of research on DR accommodation in the system. It includes mainly two types of approaches viz. price-based demand response (PBDR) and incentive-based demand response (IBDR). PBDR inflicts dynamic price on customers to change their load pattern and IBDR lays out incentive to customers during peak hours for curtailing their load~\cite{Aalami2010a}. \\
DR is one of the kinds which is driven by the utilities and executed by customers. Thus, the customers’ behaviour is essential to understand DR. The customer’s behavioral analysis is extensively studied in micro-economics, a part of economics~\cite{OtaniYoshihikoEl-Hodiri1987}. Many of its applications are also being adopted in the applications of DR. In this, customers’ preference model, game theory and price elasticity model (PEM) are widely employed approaches in DR modeling as described in the literature~\cite{Kirschen2000b,Mohajeryami2015, Mohajeryami2016b, Shinde2018b, Deng2015a}. Where, utility functions are utilized for defining customer priority, game theory for decision making in the competitive environment and PEM for assessing the load deviation with price variation. This article covers the customer’s sensitivity to price variation using PEM for studying load pattern variation.  \\
\begin{table}[!h]   
	\begin{framed}
		\mbox{}
		\nomenclature[I]{$n, N$}{Index and set of load bus.}
		\nomenclature[I]{$t,\tau, T$}{Index and set of time.}
		\nomenclature[I]{$cl, CL$}{Index and set of customer classes.}
		\nomenclature[I]{$i, I$}{Index and set of customer number.}
		\nomenclature[I]{$T_{P/OP/V}$}{Set of peak, off-peak and valley hours.}
		\nomenclature[P]{$\rho_{t}$}{Utility selling real time price at time $t$.}
		\nomenclature[P]{$\rho_{t}^{cl}$}{Real time price offered to class $cl$ at time $t$.}
		\nomenclature[P]{$\kappa^{cl}$}{Price coefficient for charging/subsidy over different classes.}
		\nomenclature[P]{$\alpha_{i}^{cl}$}{Factor of  of  load curtailment. }
		\nomenclature[P]{$D_{i}^{cl}(t)$}{Demand after DR of $i$th customer of class $cl$ at time $t$.}
		\nomenclature[P]{$D_{i,o}^{cl}(t)$}{Demand before DR of $i$th customer of class $cl$ at time $t$.}
		\nomenclature[P]{$\Delta D_{i}^{cl}(t)$}{Change in demand of $i$th customer of class $cl$ at time $t$.}
		\nomenclature[P]{$W$}{Weiner/Brownian variable.}
		\nomenclature[P]{$\lambda_{i}^{cl}$}{Lagrange multiplier of $i$th customer of class $cl$.}
	   	\nomenclature[P]{$\mu, \sigma$}{Mean and standard deviation .}
	   	\nomenclature[P]{$LF_{i}^{cl}$}{Load factor of $i$th customer of class $cl$.}
	   	\nomenclature[P]{$a_{t}^{cl}, b_{t}^{cl}$}{Lower and upper bound of truncated normal distribution.}
	   	\nomenclature[F]{$S(.)$}{Net benefit/Social welfare function.}
	   	\nomenclature[F]{$B(.)$}{Customer benefit function.}
	   	\nomenclature[F]{$L(.)$}{Lagrange function.}
	   	\nomenclature[F]{$N(.)$}{Normal distibution.}
	   	\nomenclature[F]{$\Phi(.)$}{Cumulative normal distribution.}
	   	\nomenclature[F]{$ \Phi^{-1}(.)$}{ Inverse cumulative normal distribution.}
	   	\nomenclature[S]{$\underline{(.)}, \overline{(.)}$}{Upper and lower bound of (.)}
	   	\nomenclature[S]{$({\hat{.}})$}{Mean value of (.).}
		\printnomenclature
	\end{framed}
	
\end{table}
An early approach of PEM application in DR is studied in~\cite{Kirschen2000b} where authors described the customer’s reactions to price change using PEM. The authors explain that self and cross-elasticity of demand can be utilized for setting up price in the centralized electricity market. A price responsive economic load model is investigated in~\cite{Aalami2010a} using PEM under the various PBDR programs. The analysis reveals that in most programs, DR achieve better peak reduction and improved load factor. But from the economic point of view, utility revenues are increased at the cost of customers’ bills. A detailed study is carried out on different non-linear functions such as logarithmic, exponential and power structure based on the concept of customer benefit and price elasticity for measuring the better DR under the dynamic price~\cite{aalami2015evaluation}. The results show that under small elasticity and small price deviation all models give nearly same response. Among all, power structure function is most conservative model for the reliable operation of the systems. A composite weighted DR function amalgamation of linear and non-linear functions is utilized to assess the customer’s DR in dynamic price~\cite{Yousefi2011}. In addition, demand function has been modeled on the basis of historical data of price and demand. In~\cite{Safdarian2014a,Safdarian2015} authors have utilized the PEM in short and medium-term decision-making models via real time price (RTP) and time-of-use (TOU) model to measure the impact of DR on distribution company’s profit. A similar framework is also considered in~\cite{Gutierrez-Alcaraz2016a, Zeng2014, Baboli2012a}, where the authors evaluated elastic/flexible demand under the different PBDR programs. However, in most of these studies~\cite{Aalami2010a, Kirschen2000b, Gutierrez-Alcaraz2016a, Zeng2014, Baboli2012a}, DR is evaluated at utility level using an aggregated load approach with the static value of elasticity.    \\
 Though, a need of disaggregated and distinct elasticity approach is also envisioned in the literature. A disaggregated load profiling approach and different elasticity value based on load sector and appliances is considered in~\cite{roscoe2010supporting} to observe the flexible demand under the real time environment for wind penetration. The author stresses on the development of intelligent demand forecasting algorithm to account time and weather conditions. Moreover, this study discussed the need of different electricity price patterns to the different customers via wide range of tariff for better participation level in DR. In Ref.~\cite{Zhao2013multi}, PEM is modeled stochastically considering the various scenarios with a predefined range of elasticity value for measuring the varying response of customers under the dynamic price. In Ref.~\cite{Gomez2012learning}, the authors estimated individual elasticity of customers through sparse construction method using linear regression after the change in the price. To better illustrate elasticity, a microscopic approach for residential customers is observed in~\cite{Asadinejad2018} by considering an appliance-wise elasticity based on survey under IBDR. The results show that this disaggregated approach gives better visualization of flexible load in DR. In Ref.~\cite{Venkatesan2011}, the distinctive elasticity coefficients are assumed for all states, further these coefficients are elaborated distinctively to describe the customers’ possible reaction in DR. However, there is no meticulous analytical model for assuming these values. Similar coherence is also deliberated in~\cite{Dadkhah2017network}, where the authors stated that the price increment/decrement during distinct time periods cannot give same DR due to customer’s non-linear behaviour i.e. elasticity value must be different to show the relative effect. The study has assumed low elasticity value for low price periods and high elasticity for high price periods. A better dynamic elasticity extraction approach is suggested in~\cite{Yousefi2011}, where demand functions are correlated with the price on the basis of fitting of historical data of price and demand. This gives the distinct elasticity value at each hour as opposite to fixed value approach considered in the previous works. \\
 In the light of above literature review, it can be assessed that PEM is an effective approach to observe the customer’s reaction to DR. However, the application of PEM for DR evaluation does not fully apprehend its attributes of adaptability and adjustability. These two attributes depend upon the customers’ flexibility, and are exhibited through elasticity. Elasticity is obtained as the relative variation in the demand to the relative change in the price. Though, this relative variation with increment/decrement does not follow same elasticity pattern due to non-linear behaviour of customers~\cite{Kirtley2008}. This will give partial adaptiveness in DR. In addition, the assumption of static elasticity lowers the interrelation effect existing between price and demand over the multi-state framework. This makes sense to use of dynamic elasticity for replicating adaptability in DR, pooled between peak hours and off-peak/valley hours. Therefore, an adaptive economic DR framework is proposed using PEM and realization of elasticity is established through a deterministic and a stochastic approach. Here in, deterministic approach, the elasticity value is being made dynamic to interrelate elasticity of peak and off-peak/valley hours. This dynamic elasticity capsulizes the attributes of self and cross-elasticity. In stochastic approach, elasticity is modeled as the stochastic process to exhibit the uncertain behaviour of customers in DR. A geometric Brownian motion (GBM) is utilized to model the cross-elasticity to emulate flexibility evolution over the time. The proposed stochastic model incorporates intertemporal constraints to imitate load recovery for shiftable/flexible loads, which was not explicitly modeled in PEM. It presents a realistic load recovery with intertemporal structure of load shift. Besides, a disaggregated load approach for DR valuation at customer level is adopted in contrast to aggregated load as considered in the literature. For this, a bottom-up approach is considered to accommodate different classes and their diversified customers. 
 The main contributions of the paper are summarized as follows:
 \begin{itemize}
 	\item An adaptive economic DR framework using PEM is developed to reflect the factual DR behaviour. 
 	\item To innovate an adaptive DR through a dynamic elasticity using a deterministic approach. The proposed approach establishes an interlink between peak and off-peak/valley hours’ elasticity to realize complete load recovery for the curtailable/shiftable load. 
 	\item A stochastic approach is envisioned to model the uncertain behaviour of customers in DR using PEM. For this, a stochastic process, GBM is utilized to imitate load recovery in DR with better intertemporal time constraint. 
 	\item	A class wise load profiling to assess each customer’s response in DR.
 	
 \end{itemize}
 The remaining paper is planned as follows Section 2 describes proposed methodology and the motivation behind it. Section 3 describes the load profiling. Result and comparison analysis is performed in the Section 4. Finally, conclusion of paper is presented in Section 5.\\
 
\section{Proposed adaptive economic DR framework  }
The customer’s DR under the ambit of dynamic price is measured as the difference of customer benefit and their payment to the utility~. This function is termed as net benefit function or social welfare function. The customers benefit and its payment vary with the demand capacity. Therefore, a disaggregated approach is a better realization to observe the distinct response as suggested in~\cite{Schweppe1988}. It is also assumed that all the customers will behave rationally in DR for the ease of exposition. On the basis of this, the net benefit function for each customer of different class can be expressed as follows: \\
\begin{equation}\label{eq.1}
S(D_i^{cl}(t)) = B(D_i^{cl}(t)) - D_i^{cl}(t){\rho ^{cl}}(t)
\end{equation}
When customers participate in DR, they can optimistically expect: i) to get the benefit from participating in DR program and ii) to maintain the overall consumption of energy same. These two expectations can only be coherent, when customers are adaptive and adjustable to DR program. Thus, to make customer’s adaptable and adjustable to the DR program, the problem is formulated as a constraint optimization to redistribute the demand optimally. This allows the customer to keep its overall consumption same before DR (BDR) and after DR (ADR). Based on this assumption, the following constraint is imposed on net benefit function and is expressed as follows: \\
\begin{equation}\label{eq.2}
	\centering
\psi _i^{cl} = \sum\limits_{t \in {T_P}} {\Delta D_i^{cl}(t)}  -\sum\limits_{t \in \{ {T_{OP}} \cup {\rm{ }}{T_V}\} } {\Delta D_i^{cl}(t)}  = 0
\end{equation}
This transforms the customer benefit function into a constraint optimization problem. Such problem can be effectively optimized using a Langrage function, which add the constraint into the objective function and multiply the constraint by a coefficient, called as Lagrange multiplier~\cite{DimitriP.Bertsekas1982} . The Lagrange function for the proposed formulation can be expressed as follows: \\
 \begin{equation}\label{eq.3}
 	\centering
 L(D_i^{cl}(t)) = S(D_i^{cl}(t)) + \lambda _i^{cl}\psi _i^{cl}
 \end{equation}
  \begin{equation}\label{eq.4}
  	\centering
 L(D_i^{cl}(t)) = S(D_i^{cl}(t)) + \lambda _i^{cl}(\sum\limits_{t \in {T_P}} {\Delta D_i^{cl}(t) - } \sum\limits_{t \in \{ {T_{OP}} \cup {\rm{ }}{T_V}\} } {\Delta D_i^{cl}(t)} )
 \end{equation}
 The second term in (\ref{eq.4}) represents the difference of demand exchanged between the peak and valley/off-peak hours. It is to be noted that the peak, valley and off-peak hours are pre-determined and distinct for the sake of simplicity. Further, substituting (\ref{eq.1}) and (\ref{eq.2}) into (\ref{eq.3}), we get  \\
   \begin{equation}\label{eq.5}
   	\centering
L(D_i^{cl}(t)) = B(D_i^{cl}(t)) - D_i^{cl}(t){\rho ^{cl}}(t) + \lambda _i^{cl}(\sum\limits_{t \in {T_P}} {\Delta D_i^{cl}(t) - } \sum\limits_{t \in \{ {T_{OP}} \cup {\rm{ }}{T_V}\} } {\Delta D_i^{cl}(t)} )
 \end{equation}
    \begin{equation}\label{eq.6}
    	\centering
	L(D_i^{cl}(t)) = B(D_i^{cl}(t)) - D_i^{cl}(t){\rho ^{cl}}(t) + \lambda _i^{cl}\left\{ {\sum\limits_{t \in {T_P}} {(D_{i,o}^{cl}(t) - D_i^{cl}(t)) - } \sum\limits_{t \in \{ {T_{OP}} \cup {\rm{ }}{T_V}\} } {(D_i^{cl}(t)}  - D_{i,o}^{cl}(t))} \right\}
\end{equation}
The optimal condition of the Lagrange function can be defined for the different time periods by taking the first derivative of (\ref{eq.6}).\\
\begin{equation}\label{eq.7}
	\centering
	\frac{{\partial L(D_i^{cl}(t))}}{{\partial D_i^{cl}(t)}} = \frac{{\partial B(D_i^{cl}(t))}}{{\partial D_i^{cl}(t)}} - {\rho ^{cl}}(t) - \lambda _i^{cl}{\rm{      }}~~~~~\forall t
\end{equation}
It can be seen from (\ref{eq.7}) that the marginal rate of benefit function varies with the price of that hour. \\
  \begin{equation}\label{eq.8}
  	\centering
	\frac{{\partial B(D_i^{cl}(t))}}{{\partial D_i^{cl}(t)}} = {\rho ^{cl}}(t) + \lambda _i^{cl}{\rm{      }}~~~~~\forall t 
\end{equation}
The second derivative of (\ref{eq.8}) is defined as following. \\
 \begin{equation}\label{eq.9}
	\centering
	\frac{{\partial {B^2}(D_i^{cl}(t))}}{{\partial D_i^{cl}{{(t)}^2}}} = \frac{{\partial {\rho ^{cl}}(t)}}{{\partial D_i^{cl}(t)}}
\end{equation}
The customer benefit function most often written in the quadratic form using the Taylor series expansion is expressed as follows:~\cite{Schweppe1988}. \\
\begin{equation}\label{eq.10}
\centering
B(D_i^{cl}(t)) = B(D_i^{cl}(t)) + {B^{'}}(D_i^{cl}(t))(D_i^{cl}(t) - D_{i,o}^{cl}(t)) + {B^{''}}(D_i^{cl}(t))\frac{{{{(D_i^{cl}(t)-D_{i,o}^{cl}(t))}^2}}}{2}
\end{equation}
\begin{equation}\label{eq.11}
	\centering
	B(D_i^{cl}(t)) = B(D_i^{cl}(t)) + (\rho _o^{cl}+\lambda_{i}^{cl})(t)(D_i^{cl}(t) - D_i^{cl}(t)) + \frac{{\rho _o^{cl}(t)}}{{\varepsilon _i^{cl}(t)D_{i,o}^{cl}(t)}}\frac{{{{(D_i^{cl}(t) - D_{i,o}^{cl}(t))}^2}}}{2}
\end{equation}
Differentiating (\ref{eq.11}) and equating it with (\ref{eq.8}) gives the following expression. \\
\begin{equation}\label{eq.12}
	\centering
	{\rho ^{cl}}(t) + \lambda _i^{cl}{\rm{ }} = \rho _o^{cl}(t)\left\{ {1 + \frac{{(D_i^{cl}(t) - D_{i,o}^{cl}(t))}}{{\varepsilon _i^{cl}(t)D_{i,o}^{cl}(t)}}} \right\}
\end{equation}
Rearranging the expression, the customer’s demand at any time $t$ ADR will be given by:\\ 
\begin{equation}\label{eq.13}
		\centering
	D_i^{cl}(t) = D_{i,o}^{cl}(t)\left\{ {1 + \varepsilon _i^{cl}(t)\frac{{({\rho ^{cl}}(t) - \rho _o^{cl}(t) + \lambda _i^{cl})}}{{\rho _o^{cl}(t)}}} \right\}{\rm{  }}~~~~~\forall t \in T
\end{equation}
Equation (\ref{eq.13}) represents the demand ADR using self-elasticity. Similarly, to incorporate the effect of cross-elasticity, PEM can be extended into multi periods as follows~\cite{Aalami2010a}: \\
\begin{equation}\label{eq.14}
	 	\centering
D_i^{cl}(t) = D_{i,o}^{cl}(t)\left\{ {1 + \varepsilon _i^{cl}(t)\frac{{({\rho ^{cl}}(t) - \rho _o^{cl}(t) + \lambda _i^{cl})}}{{\rho _o^{cl}(t)}} + \sum\limits_{\tau  \in T} {\varepsilon _i^{cl}(t,\tau )\frac{{({\rho ^{cl}}(\tau ) - \rho _o^{cl}(\tau ) + \lambda _i^{cl})}}{{\rho _o^{cl}(\tau )}}} } \right\}{\rm{  }}~~~~~\forall t \in T,~\forall \tau  \in T
\end{equation}
Equation (\ref{eq.14}) represents the combined effect of self and cross-elasticity on the demand. Where, self-elasticity manifests increment/decrement relative to change in the price on the same period, whereas, cross-elasticity evaluates the transverse effect of price on the cross-demand. It is mathematically defined as follows:
   \begin{equation}\label{eq.15}
	\centering
	\varepsilon (t,\tau ) = \left( {\frac{{\Delta D(t,\tau )}}{{\Delta \rho (t,\tau )}}} \right)\left( {\frac{{{\rho _o}(t,\tau )}}{{{D_o}(t,\tau )}}} \right)
\end{equation}
\begin{equation}\label{eq.16}
	\centering
	\left\{ {\begin{array}{*{20}{c}}
			{\varepsilon (t,\tau ) \le 0}&{{\rm{if }}t = \tau }\\
			{\varepsilon (t,\tau ) > 0}&{{\rm{if }}t \ne \tau }
	\end{array}} \right.
\end{equation}
Equation (\ref{eq.16}) denotes the notation of self and cross-elasticity, where self-elasticity is attributed to negative value and positive value is assigned to latter. From the electricity point of view, the magnitudes of both the elasticities vary with the customer’s reaction to price change in DR. Thus, assessment of each type of customer requires a rigorous approach, which is cumbersome task due to complex behaviour of customers. Though, with some approximate assumptions (as considered above) an analysis can be carried out to assess the impact of DR using PEM.  In this context, this study presents a deterministic and a stochastic approach to model DR using PEM. In deterministic approach, a dynamic elasticity encapsulating the features of self and cross-elasticity is proposed. Likewise, in stochastic PEM, a cross-elasticity is modeled using a geometric Brownian motion (GBM) to imitate customer behaviour under the load recovery.  \\
\subsection{Proposed deterministic PEM (DPEM) }
\begin{figure}[hbt!]
	\centering 
	\includegraphics[width=0.45\textwidth]{{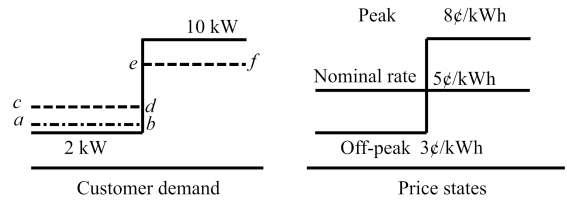}}
	\caption{Schematic diagram of two states under static and dynamic elasticity. }
	\label{DR_1}
\end{figure}

In DR, the increase or decrease of load in single time-period is achieved through switching control or continuous control of the load and is modeled by the self-elasticity. On the other hand, cross-elasticity is defined for shiftable loads. However, the interrelation between both the elasticities is merely discussed in the literature. Though, in Ref.~\cite{Kirschen2000b, Kirtley2008} the authors have touched upon the issue of optimal redistribution of demand in DR using PEM. These studies states that the optimal redistribution of load is possible when the following condition is satisfied:\\
   \begin{equation}\label{eq.17}
	\centering
	\sum\limits_{\tau  \in T} {\varepsilon (t,\tau ) = 0} {\rm{     }}~~~~~\forall t 
\end{equation}
Equation (\ref{eq.17}) indicates that whatever demand is curtailed using self-elasticity should be adjusted in the cross-time periods through the cross-elasticity. It defines the loss-less situation (i.e. no change in the consumption BDR and ADR. The aforementioned equation advocates that an interrelation should exist between self and cross-elasticity for an adaptive and adjustable DR framework. To keep this in a view, a dynamic elasticity model attributing the features of both the elasticities is envisioned.\\
The basis of the proposed dynamic elasticity is to make DR adaptive, which depends upon the customer’s flexibility and stretchiness across the periods. Therefore, a dynamic elasticity is proposed for exhibiting the customer’s flexibility from peak hours to the low-price hours. It establishes an interrelation between peak hour elasticity to off-peak/valley hours for better adaption of DR. As, it is presumed that the shifted demand during off-peak/valley hours is a reflection of curtailed peak demand. In the similar way, it can be assumed that elasticity during off-peak/valley hours will be reflection of peak hours’ elasticity. For this assumption to stand, the product of elasticity and demand at any hour is assumed to be equal to product of elasticity and demand at peak hour. This gives elasticity value relative to peak hour elasticity defined as follows: 
\begin{equation}\label{eq.18}
	\centering
	\varepsilon _i^{cl}(t)D_{i,o}^{cl}(t) = \varepsilon _i^{cl}({t_{peak}})D_{i,o}^{cl}({t_{peak}}){\rm{   }}~~~~~\forall t \in T\backslash\left\{{t_{peak}}\right\}
\end{equation}
The following relationship is defined in such a way that if demand and expected elasticity (i.e. possible peak load curtailment) at peak hour are known, then elasticity of other time states can be quantified in reference to peak hour for achieving the presumed assumption. The hypothesized relationship can be explained by considering two states (off and on peak) example for the clarity as depicted in Fig.~\ref{DR_1} \\
 \\
Let the nominal electricity price is 5 ¢/kWh and dynamic price is 3 ¢/kWh and 8 ¢/kWh at off-peak and peak hour, respectively. The corresponding demand before DR is 2 kW and 10 kW at nominal price. Now, if DR in initiated with price variation. The customer will respond to price by varying the demand evaluated using price elasticity. If it is assumed that the peak hour elasticity is 0.2. Then curtailed load demand at peak hour will be 1.2 kW marked by line $ef$ calculated using (\ref{eq.15}). While at off-peak hour the demand is raised by 0.24 kW defined by line $ab$ for the same elasticity value. This reflects partial DR. However, through the proposed dynamic elasticity approach, the elasticity value at off-peak will be equal to 1 as evaluated using (\ref{eq.18}) relative to peak elasticity. In this case the load demand is increased by 1.2 kW denoted by line $cd$, which is equal to curtailed load demand.  \\
The determination of peak hour is another complex task due to different load pattern of customer classes. For this, two approaches are suggested. In first, the peak hour is that where customers’ maximum demand occurs. At this hour, elasticity value will be assumed suitably and other hours’ elasticity will be calculated relatively to peak hour.  \\
\begin{equation}\label{eq.19}
	\centering
	\varepsilon _i^{cl}(t) = \varepsilon _i^{cl}({t_{peak}})\left[ {\frac{{\max (D_{i,o}^{cl})}}{{D_{i,o}^{cl}(t)}}} \right]{\rm{  }}~~~~~\forall t \in T\backslash\left\{{t_{peak}}\right\}
\end{equation}
In second approach, peak hour is defined on the basis of peak price. Since,  the customer will have highest sensitivity at peak price. Thus, it is chosen as peak hour. Then, at this hour the permissible demand which may be possibly curtailed for DR is assumed to be the fraction of the demand at peak price. Based on this the elasticity value is evaluated at peak hour. The other hour elasticity is evaluated using (\ref{eq.19}). \\
\begin{equation}\label{eq.20}
	\centering
\begin{array}{l}
	\Delta D_i^{cl}(t){\left. { = \alpha _{_i}^{cl}D_{i,o}^{cl}(t)} \right|_{t = {t_{peak}} \buildrel \wedge \over = \rho _{peak}^{cl} = \max ({\rho ^{cl}})}}\\
	\varepsilon _i^{cl}({t_{peak}}) = {\left. {\left( {\frac{{\Delta D_{i,o}^{cl}(t)}}{{\Delta {\rho ^{cl}}(t)}}} \right)\left( {\frac{{\rho _o^{cl}}}{{D_{i,o}^{cl}(t)}}} \right)} \right|_{t = {t_{peak}} \buildrel \wedge \over = \rho _{peak}^{c;} = \max ({\rho ^{cl}})}}\\
	\varepsilon _i^{cl}(t) = \varepsilon _i^{cl}({t_{peak}})\left[ {\frac{{D_i^{cl}({t_{peak}})}}{{D_i^{cl}(t)}}} \right]{\rm{     }}~~~~~\forall t \in T\backslash\left\{{t_{peak}}\right\}
\end{array}
\end{equation}
  It is to be noted that $\alpha_{i}^{cl}$ is the flexible demand of total demand of a customer. Thus, it varies with the customer’s activity usage and classes. Once the elasticity at every hour is defined using (\ref{eq.19}). The final expression of demand after DR is given by following expression.\\
\begin{equation}\label{eq.21}
	\centering
	D_i^{cl}(t) = D_{i,o}^{cl}(t) + \varepsilon _i^{cl}({t_{peak}})\left[ {\frac{{\max (D_{i,o}^{cl})}}{{D_{i,o}^{cl}(t)}}} \right]{\rm{ }}D_{i,o}^{cl}(t)\left( {\frac{{{\rho ^{cl}}(t) - \rho _o^{cl}(t) + \lambda _i^{cl}}}{{\rho _o^{cl}(t)}}} \right){\rm{  }}~~~~~\forall t \in T,\forall i \in I
\end{equation}
The one of the main feature of the proposed dynamic elasticity model that it is a time-variant model, which eliminates the drawbacks of the existing model, where elasticity is treated as static value as considered in the literature ~\cite{Kirschen2000b}. In addition, it amalgamates the attribute of self and cross-elasticity in the dynamic elasticity. It can be understood through the distinct trait of elasticity as discussed earlier. Since, self-elasticity will be relatively low during off-peak hours due to the small relative change in price~\cite{Dadkhah2017network}, which makes demand increase small. Thus, the cross-elasticity should cause a significant increase in the demand during off-peak hours. It reflects the adaptiveness of the customers to shift the demand on low price hours on the account of high price during peak hours. \\
\subsection{Proposed stochastic PEM (SPEM) modeling using geometric Brownian motion }
In this section, the elasticity is modeled through the stochastic approach. As, it is discussed above that self and cross-elasticity should be interrelated to make DR adaptive. It can be better elaborated using PEM as shown in (\ref{eq.22}).\\
\begin{equation}\label{eq.22}
	\centering
\left[ {\begin{array}{*{20}{c}}
		{\Delta D(1)}\\
		\begin{array}{l}
			\Delta D(2)\\
			\vdots 
		\end{array}\\
		{\Delta D(T)}
\end{array}} \right] = \left[ {\begin{array}{*{20}{c}}
		{\varepsilon (1,1)}&{\varepsilon (1,2)}& \cdots &{\varepsilon (1,T)}\\
		{\varepsilon (2,1)}&{\varepsilon (2,2)}& \cdots &{\varepsilon (2,T)}\\
		\vdots & \vdots & \ddots & \vdots \\
		{\varepsilon (T,1)}&{\varepsilon (T,2)}& \cdots &{\varepsilon (T,T)}
\end{array}} \right]\left[ {\begin{array}{*{20}{c}}
		{\Delta \rho (1)}\\
		\begin{array}{l}
			\Delta \rho (2)\\
			\vdots 
		\end{array}\\
		{\Delta \rho (T)}
\end{array}} \right]
\end{equation}
Equation (\ref{eq.22}) is expressed in per unit value. It can be observed from (\ref{eq.22}) that change in the demand at any time $t$ is the result due to that time period and other cross-periods. This correlation between both the elasticities varies with the measure of adaptability and adjustability of DR framework. This adaptability and adjustability of the demand will be complete or lossless, when the overall change in the demand over the time horizon will be zero~\cite{Jiang2019}.
\begin{equation}\label{eq.23}
	\centering
	\sum\limits_{t \in T} {\Delta D(t)}  = \sum\limits_{t \in T} {\sum\limits_{\tau  \in T} {\varepsilon (t,\tau )\Delta \rho (t){D_o}(t) = 0} } {\rm{     }}~~~~~\forall t,~~\forall \tau 
\end{equation}
Equation (\ref{eq.23}) is a complete form of (\ref{eq.17}) suggested by~\cite{Jiang2019}, which states that the overall consumption BDR and ADR should be zero over the time horizon. It indicates that price-responsive demand will be redistributed optimally in the cross-time periods. If it does not satisfy that condition, then following relation will hold: 
\begin{equation}\label{eq.24}
	\centering
	\sum\limits_{t \in T} {\Delta D(t) < 0} 
\end{equation}
Equation (\ref{eq.24}) indicates the reduction in the demand after DR. From the DR perspective, the cross-elasticity is a positive constant, because it gives rise to the cross-demand on the account of high price on another time period. This indicates that the cross-elasticity reflects the curtailed/shifted demand in the cross-periods as sort of load recovery. Though, the recovery of the load in the cross-periods governs by the flexibility of intertemporal time constraint. The intertemporal time constraint indicates that the customer can recover its curtailed demand to an extent over the cross-periods~\cite{Goransson2014, Zerrahn2015}. In Ref.~\cite{Goransson2014}, the authors have applied linear intertemporal constraints to mimic the flexible band of time for shiftable loads. Further, energy balance constraint is applied to ensure no loss of load over the time frame considered.  This concept of load recovery is also rendered in~\cite{Nguyen2012modeling} using the utility theory and in~\cite{Tulabing2016modeling} using flexibility band for different types of loads. In~\cite{Nguyen2012modeling}, the authors explained that as the customer’ delay or curtail its demand during high price time period, an exponentially deceasing customer’s utility will follow across the other time periods. This diminishing utility gives shrinking demand with the propagation of time. Similar analogy also can be emulated for the cross-elasticity value, which is responsible for the load recovery in PEM. Since, quantification of deterministic value of cross-elasticity over the time is a cumbersome task due to uncertainty associated with the customer activity usage and price. Therefore, a time varying value of cross-elasticity can be perceived as the stochastic process. In this paper, a geometric Brownian motion (GBM) is employed to estimate the time-varying value of cross-elasticity of PEM over the time~\cite{Thompson1985, Etheridge2002}. It is an extension of Brownian motion and exclusively utilized for predicting the stock price as price being positive~\cite{Etheridge2002}. Likewise, cross-elasticity is always positive from DR point of view. Thus, it is suitable for defining the randomness in the cross-elasticity. A stochastic process ${\left\{ {\varepsilon (t)} \right\}_{t \ge 0}}$ is said to be GBM, if it satisfies following stochastic differential equation (SDE) expressed as follows:\\
\begin{equation}\label{eq.25}
	\centering
	d\varepsilon (t) = \mu \varepsilon (t)dt + \sigma \varepsilon (t)dW(t)
\end{equation}
where, $\epsilon$ is defined as the stochastic variable with drift parameter $\mu$ as a propagation of the stochastic variable and volatility parameter $\sigma$ as the variation in the propagation. $W(t)$ is called as the Weiner/Brownian variable. The Weiner variable satisfies the following conditions 1) $W({t_0}) = 0$   and 2) $W({t_k}) = W({t_{k - 1}}) + N(0,{t_k} - {t_{k - 1}})$ where, $N$ being a normal distribution with zero-mean and variance equal to $\left( {{t_k} - {t_{k - 1}}} \right)$ . If ${t_0} < {t_1} < .... < {t_{k - 1}} < {t_k} < T$  are the successive interval over the time horizon, then increment $W({t_k}) - W({t_{k - 1}}) \sim N(0,{t_k} - {t_{k - 1}})$  are stationary and independent normally distributed random variable. On solving (\ref{eq.25}) the following expression is obtained. \\
\begin{equation}\label{eq.26}
	\centering
	\varepsilon (t) = {\varepsilon _0}\exp \left[ {\left( {\mu  - \frac{{{\sigma ^2}}}{2}} \right)t + \sigma B(t)} \right]
\end{equation}
Where, $\epsilon_{0}$ is the initial value at $t=0$. The generalize expression of GBM is given by (\ref{eq.27}).\\
\begin{equation}\label{eq.27}
	\centering
	\varepsilon ({t_{k + 1}}) = \varepsilon ({t_k})\exp \left[ {\left( {\mu  - \frac{{{\sigma ^2}}}{2}} \right)({t_{k + 1}} - {t_k}) + \sigma B({t_{k + 1}} - {t_k})} \right]
\end{equation}
Putting the value of Weiner variable in (\ref{eq.27}), gives the complete expression as follows:\\
\begin{equation}\label{eq.28}
	\centering
	\varepsilon ({t_{k + 1}}) = \varepsilon ({t_k})\exp \left[ {\left( {\mu  - \frac{{{\sigma ^2}}}{2}} \right)\Delta {t_k} + \sigma \sqrt {\Delta {t_k}} N(0,1)} \right]
\end{equation}
The solution of (\ref{eq.25}) is obtained using Ito’s formula~\cite{Etheridge2002, Lyuu2012} as expressed in (\ref{eq.28}). This equation represents the time- based estimation of cross-elasticity in the exponential form. Here, if ${t_0} < {t_1} < .... < {t_{k - 1}} < {t_k} < T$  are successive time points, then successive ratios are independent random variable in case of GBM as follows:\\
\begin{equation}\label{eq.29}
	\centering
	\frac{{\varepsilon ({t_1})}}{{\varepsilon ({t_0})}},\frac{{\varepsilon ({t_2})}}{{\varepsilon ({t_1})}},....,\frac{{\varepsilon ({t_k})}}{{\varepsilon ({t_{k - 1}})}}
\end{equation}
The solution of GBM characterize the stochastic process using the drift and volatility parameter. Thus, the estimation of cross-elasticity is attributed these two parameters, which can be evaluated through the historical data. It reveals that the value of cross-elasticity varies with the drift parameter $\left( {\mu  - {\raise0.5ex\hbox{$\scriptstyle {{\sigma ^2}}$}
		\kern-0.1em/\kern-0.15em
		\lower0.25ex\hbox{$\scriptstyle 2$}}} \right)$  and variance $\sigma^{2}$ . Further, the relationship between drift and volatilely defines the movement of the stochastic process. If $\left( {\mu  \le {\raise0.5ex\hbox{$\scriptstyle {{\sigma^2}}$}
		\kern-0.1em/\kern-0.15em
		\lower0.25ex\hbox{$\scriptstyle 2$}}} \right)$  , then stochastic process decay exponentially in the successive time intervals, which is equivalent to load recovery with non-linear behaviour. Conversely, for $\left( {\mu  > {\raise0.5ex\hbox{$\scriptstyle {{\sigma^2}}$}
		\kern-0.1em/\kern-0.15em
		\lower0.25ex\hbox{$\scriptstyle 2$}}} \right)$, stochastic process grow exponentially, which will give rise to rebound effect (i.e. when a new peak demand higher than base peak is emerged in post DR period) in DR~\cite{A.Greening2000}. These two conditions suggest that proposed GBM for cross-elasticity estimation can model the impact of load recovery in DR, effectively. However, this study focuses only on the modeling of load recovery for the former circumstance. \\

     \section{Load profiles}
\begin{figure}[hbt!]
 	\centering 
 	\includegraphics[width=0.45\textwidth]{{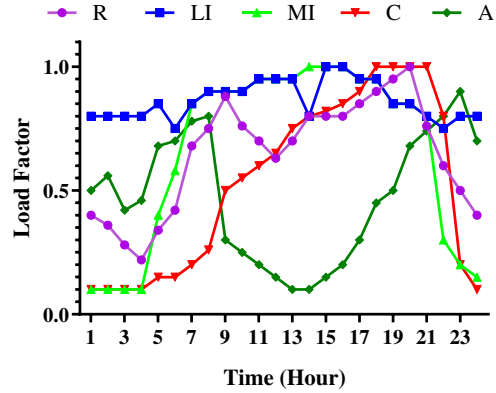}}
 	\caption{Load factor pattern of different types of customers classes.}
 	\label{Loadpattern}
 \end{figure}

     Assessing the DR at the customer level requires the knowledge of their electrical-based activity. These activity usages vary within the classes due to the various indigenous and exogenous factors and also across the different classes. This give rise to the diverse load variations within the customer class, which will also reflect on DR. Therefore, diversified load profiles are synthesized to emulate diversified DR at each individual. For this, an aggregated load pattern using load factor for each class is utilized as shown in Fig.~\ref{Loadpattern}~\cite{Kanwar2015}. Then, the variation in each class at each hour is modeled using coefficient of variation (COV). Based on the COV, the standard deviation at each hour is calculated as \\
     \begin{equation}\label{eq.30}
     	\centering
     	\sigma _t^{cl} = {\rm{COV}}_t^{cl} \times \mu _t^{cl}
     \end{equation}
     Using the mean and the standard deviation, the permissible interval of load factor variation within the same class is defined as \\
     \begin{equation}\label{eq.31}
     	\centering
     	\begin{array}{c}
     		LF_i^{cl}(t) \in [a_t^{cl},b_t^{cl}] \in [\mu _t^{cl} - \sigma _t^{cl},\mu _t^{cl} + \sigma _t^{cl}]\\
     		\in [\mu _t^{cl} - {\rm{COV}}_t^{cl} \times \mu _t^{cl},\mu _t^{cl} + {\rm{COV}}_t^{cl} \times \mu _t^{cl}]
     	\end{array}
     \end{equation}
     Equation (\ref{eq.31}) defines the possible range of variation of load factor for the considered period. To illustrate the load patterns of customers, a load factor is randomly sampled using a normal distribution. Since, normal distribution generates the variable with the range $(-\infty, \infty)$. Thus, to restrict the variables drawn from a normal distribution within a permissible definite range, a truncated normal distribution is employed. The probability distribution function (PDF) of the truncated normal distribution for each class of customers is expressed as~\cite{Burkardt2014}: \\
     \begin{equation}\label{eq.32}
     	\centering
     	\psi ( \mu _t^{cl}, \sigma _t^{cl},a_t^{cl},b_t^{cl};LF_i^{cl}(t)) = \left\{ {\begin{array}{*{20}{c}}
     			0\\
     			{\frac{{\phi ( \mu _t^{cl}, \sigma _t^{cl};LF_i^{cl}(t))}}{{\Phi (\mu _t^{cl}, \sigma _t^{cl};b_t^{cl}) - \Phi ( \mu _t^{cl}, \sigma _t^{cl};a_t^{cl})}}}\\
     			0
     	\end{array}} \right.\begin{array}{*{20}{c}}
     		{{\rm{if }}LF_i^{cl}(t) < a_t^{cl}}\\
     		{{\rm{         if }}LF_i^{cl}(t) \le LF_i^{cl}(t) \le b_t^{cl}}\\
     		{{\rm{if }}LF_i^{cl}(t) > b_t^{cl}}
     	\end{array}
     \end{equation}
     To generate a variable from truncated normal distribution the following approach is employed\\
     \begin{equation}\label{eq.33}
     	\centering
     \begin{array}{c}
     	{\rm{repeat:}}\\
     	x = rand()\\
     	LF_i^{cl}(t) = {\Phi ^{ - 1}}(\mu _t^{cl},{( \sigma _t^{cl})^2},x)\\
     	{\rm{until (}}a_t^{cl} \le LF_i^{cl}(t) \le b_t^{cl}{\rm{)}}
     \end{array}
     \end{equation}
     where, mean and standard deviation are the values of the original normal distribution. For sampling the load factor from the truncated normal distribution, a random number is generated. Then the sampled value is evaluated using an inverse cumulative density function (cdf). The generated value of the sampled load factor is checked against the condition and if it violates then repeat the process.  Based on the calculated load factor, the load profile of the customers at each load bus is synthesized using the following expression.\\
     \begin{equation}\label{eq.34}
     	\centering
     	D_{i,o}^{cl}(t) = LF_i^{cl}(t) \times \left( {\frac{{D_{i,o}^{cl}}}{{\max (D_{i,o}^{cl})}}} \right)
     \end{equation}
     \section{Results}
\begin{figure}[!h]
    	\centering 
    	\includegraphics[width=0.5\textwidth]{{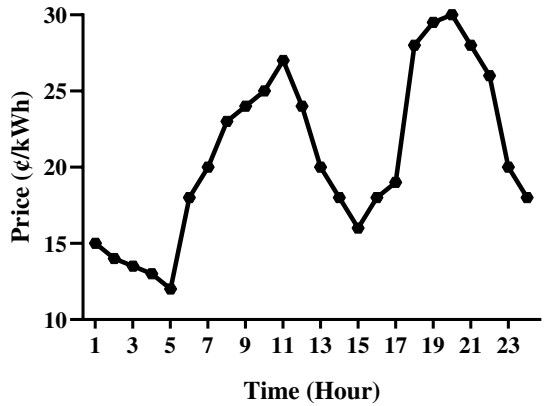}}
    	\caption{Dynamic price signal under RTP.}
    	\label{rtp}
    \end{figure}

     A case study is carried out on 33 standard distribution load bus data~\cite{Kanwar2015} to demonstrate the effectiveness of the proposed DR framework using PEM. The distribution network load buses are segmented into five different classes i.e., residential: (R), commercial: (C), agriculture: (A), large industrial: (LI) and medium industrial (MI) customers. Each load bus is independently assigned to individual class and the available load are stipulated as their demand level for the ease of exposition. The mix has the total 235 number of customers. The detail of allocated nodes, total demand, number of customers, subsidy/charging factor (S/CF) and customer demand range for each class is given in Table.~\ref{tab:custdetail}.\\
\FloatBarrier
\begin{table}[H]
     	\caption{Customer class detail}
     	\label{tab:custdetail}
     	\begin{tabular}{|l|c|c|c|c|c|}
     		\hline
     		Classes &
     		\multicolumn{1}{c|}{\begin{tabular}[c]{@{}c@{}}Allocated \\ Demand\\kW\end{tabular}} &
     		\multicolumn{1}{c|}{\begin{tabular}[c]{@{}c@{}}No. of\\  Customers\end{tabular}} &
     		\multicolumn{1}{c|}{Nodes} &
     		\multicolumn{1}{c|}{S/CF $(\kappa^{cl})$} &
     		\multicolumn{1}{c|}{\begin{tabular}[c]{@{}c@{}}Demand Range\\  {[}min, max{]}\\kW\end{tabular}} \\ \hline
     		R  & 950  & 121 & 2-10  & -0.2 & {[}2, 15{]}   \\ \hline
     		C  & 555  & 18  & 11-18 & 1    & {[}10, 25{]}  \\ \hline
     		LI & 1290 & 27  & 19-25 & 0    & {[}50, 100{]} \\ \hline
     		MI & 180  & 6   & 26-28 & 0.2  & {[}20, 45{]}  \\ \hline
     		A  & 740  & 63  & 29-33 & -0.5 & {[}8, 18{]}   \\ \hline
     	\end{tabular}
     \end{table}
    Fig.~\ref{rtp} depicts the RTP offered by the utility and is taken from Ontario Energy Board~\cite{OntarioEnergyBoard}. The price offered to the customer class is calculated using $({\rho ^{cl}}(t) = \rho (t) \times (1 + {\kappa ^{cl}}))$. It is on the assumption that each class is being offered different prices due to their demand and activity usage. The time span is partitioned into three periods namely: off-peak (0:00-8:00, 23:00-24:00), valley: (12:00-18:00) and peak: (8:00-11:00, 18:00-22:00) for the sake of simplicity. \\
         \FloatBarrier
The proposed methods are investigated class-wise and their effectiveness is evaluated in reference to standard PEM. PEM is used as a benchmark model for the comparison analysis. The class-wise elasticity value for R, LI, MI, C and A customers are -0.30, -0.43, -0.54, -0.30 and -0.23, respectively as reported in~\cite{Stewart2018}. It is worth to mention that elasticity values in case of DPEM and SPEM are evaluated using the proposed dynamic and stochastic model. Though, a peak hour elasticity for DPEM is same as the standard PEM, while in case of SPEM, the initial value of cross elasticity is assumed as the fraction of self-elasticity. It is based on the notion of low value of cross-elasticity as discussed in the literature~\cite{Kirschen2000b}. Here, the initial value of cross-elasticity is taken as $(\varepsilon (t,\tau ) \times 0.15{|_{t = \tau }})$   and subsequently evaluated through the GBM descriptive parameters (i.e. $\mu$ and $\sigma$) using (\ref{eq.28}). The values of $\mu$ and $\sigma$ are taken as 0.2 and 1.2, respectively for the analysis purpose. \\    
    \begin{figure}[!h]
	\centering 
	\includegraphics[width=0.5\textwidth]{{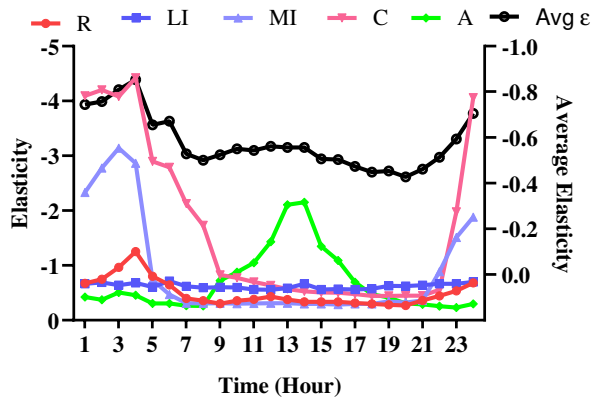}}
	\caption{An aggregated class-wise elasticity variation using proposed DPEM.}
	\label{elasticityvar}
\end{figure}\\
   An aggregated class wise elasticity variation profile using the proposed DPEM is shown in Fig.~\ref{elasticityvar}. The figure shows the average elasticity variation over the time horizon for the different classes and aggregated elasticity of all the customer classes. It can be observed from the figure that elasticity variation in R and LI class is on the nominal range. On the other hand, MI, A and C class customers exhibit wide variation in the elasticity. The rationale behind such vast elasticity variation can be explained through the customer aggregated load factor or load pattern. The figure shows that the customer class which exhibits high relative ratio between peak to valley/off-peak set forth wide elasticity variation and vice-versa. It can be substantiated from the elasticity profiles of R and LI class customers, whose relative ratios between the segmented time periods are on nominal range, while the wide-variation are reported for MI, A and C class customers due to high relative proportions. This elasticity variation indicates that load patterns with high relative ratio will need to exhibit high flexibility to achieve true DR, otherwise may face the overall reduction in their energy consumption. On the contrary, customers with low proportions among different time periods can display the load flexibility over the time horizon with the ease. \\
     \FloatBarrier
%
     \FloatBarrier

     \FloatBarrier
     \FloatBarrier
     \FloatBarrier
     \begin{figure}[!h]
     	\centering
     	\begin{subfigure}[b]{0.45\textwidth}
     		\centering
     		\includegraphics[width=\textwidth]{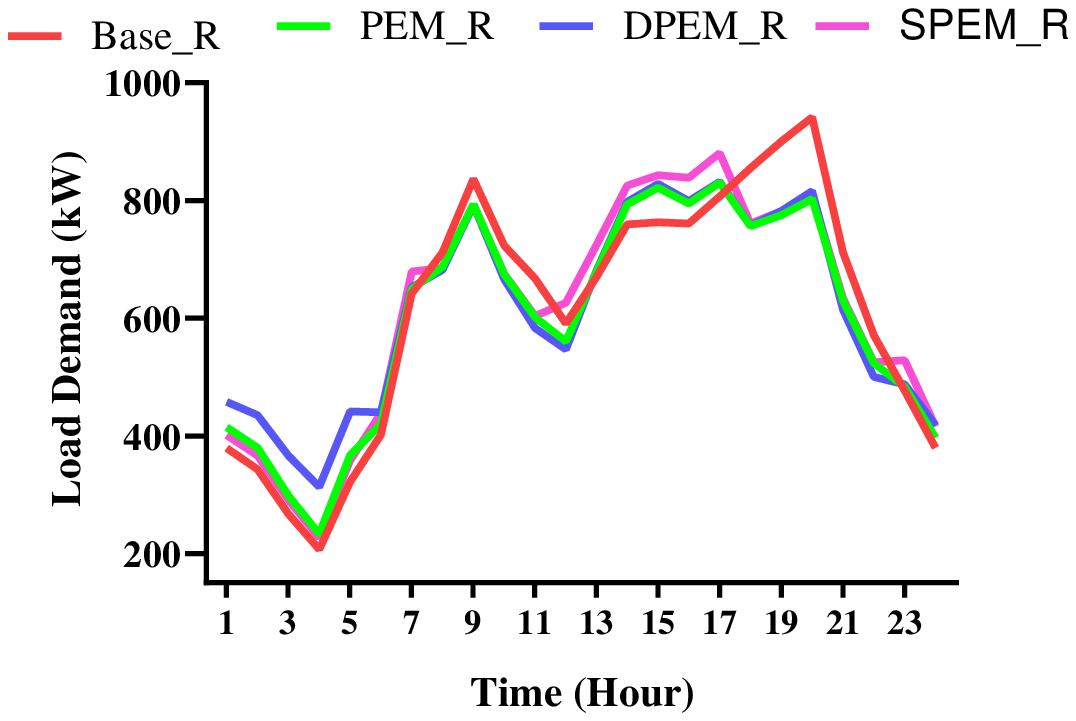}
     		\caption{}
     		\label{}
     	\end{subfigure}
     	\hfill
     	\begin{subfigure}[b]{0.45\textwidth}
     		\centering
     		\includegraphics[width=\textwidth]{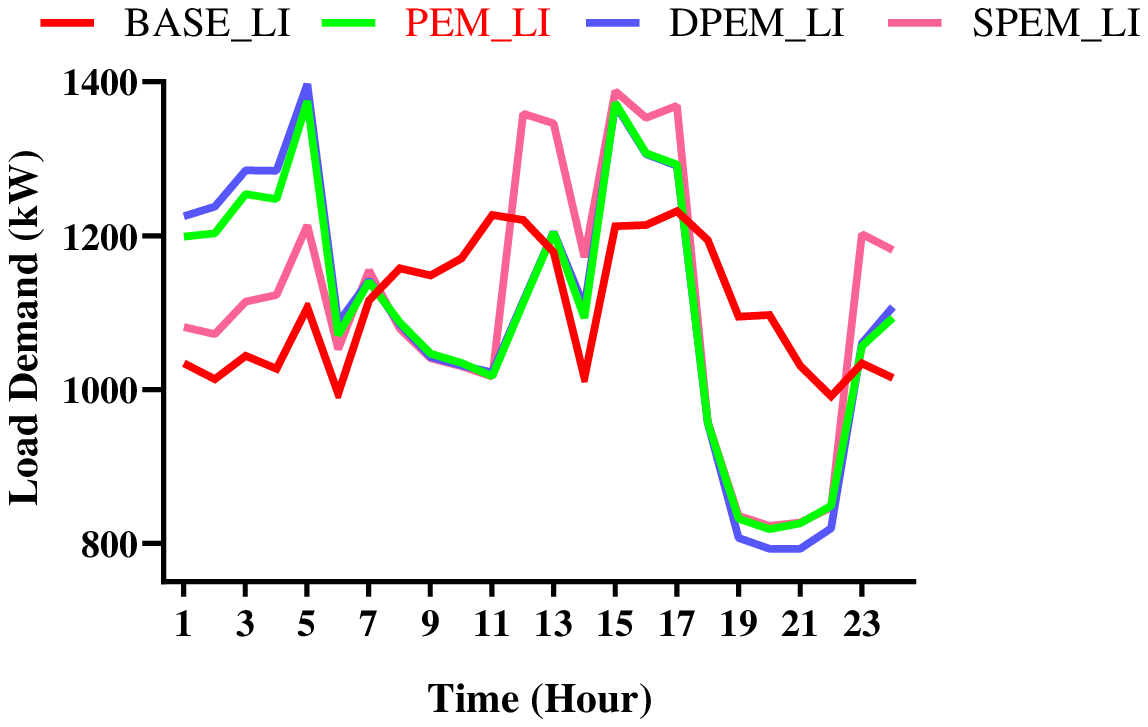}
     		\caption{}
     		\label{}
     	\end{subfigure}
     	\hfill
     	\begin{subfigure}[b]{0.45\textwidth}
     		\centering
     		\includegraphics[width=\textwidth]{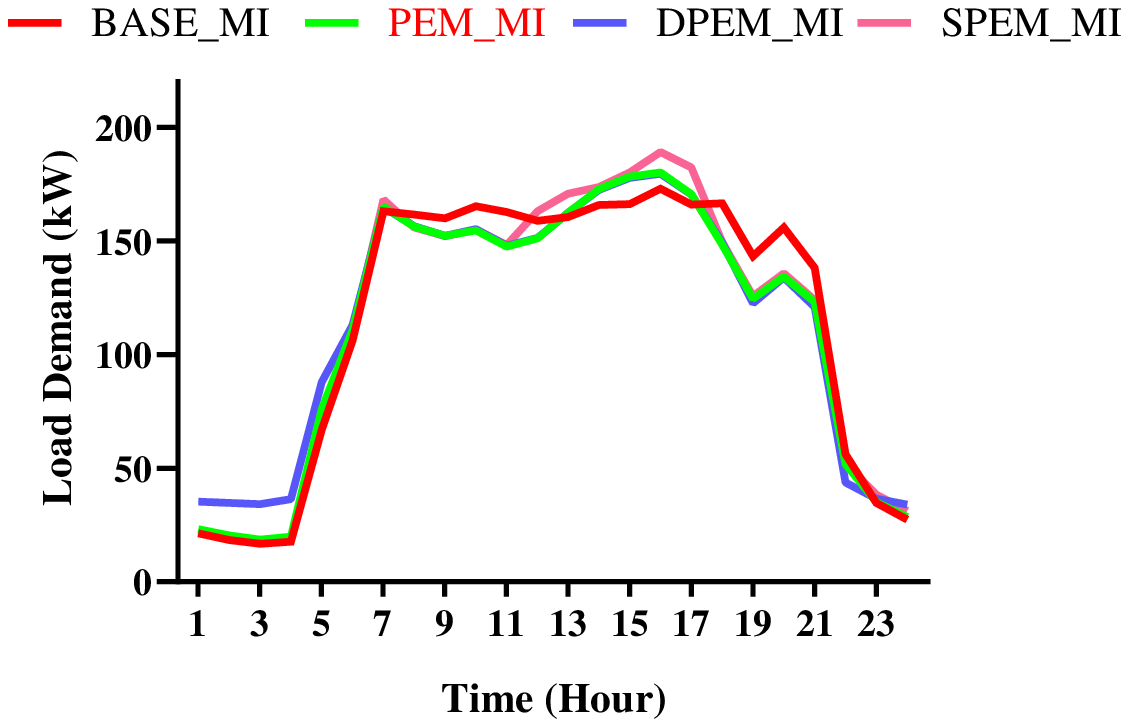}
     		\caption{}
     		\label{}
     	\end{subfigure}
     	\hfill
     	\begin{subfigure}[b]{0.45\textwidth}
     		\centering
     		\includegraphics[width=\textwidth]{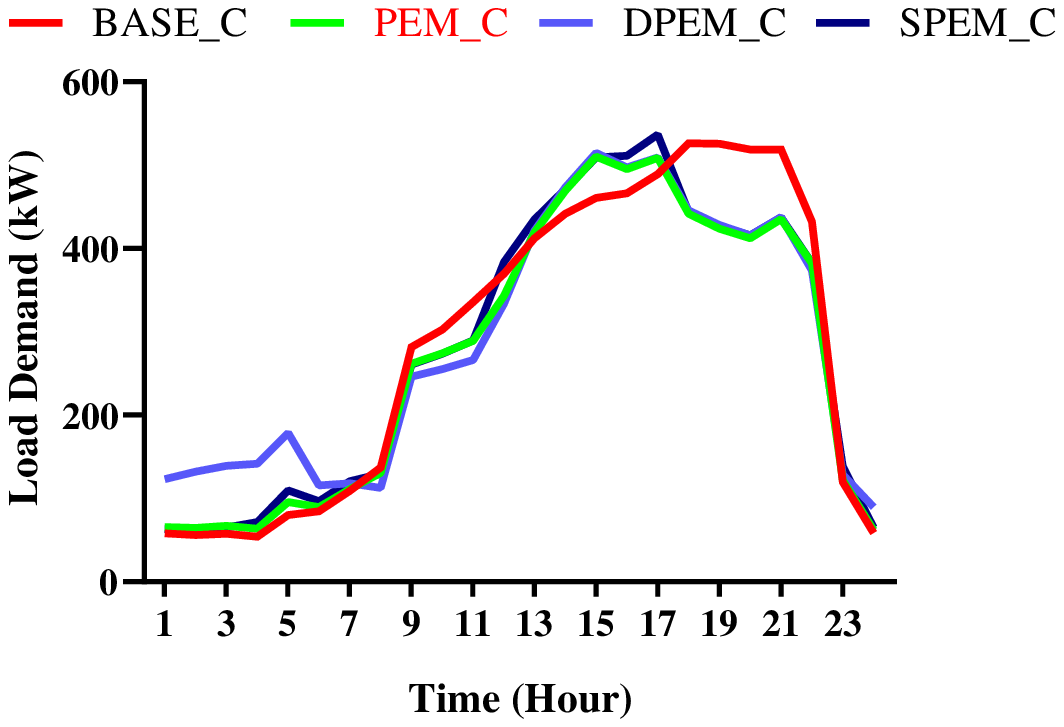}
     		\caption{}
     		\label{}
     	\end{subfigure}
     	\hfill
     	\begin{subfigure}[b]{0.45\textwidth}
     		\centering
     		\includegraphics[width=\textwidth]{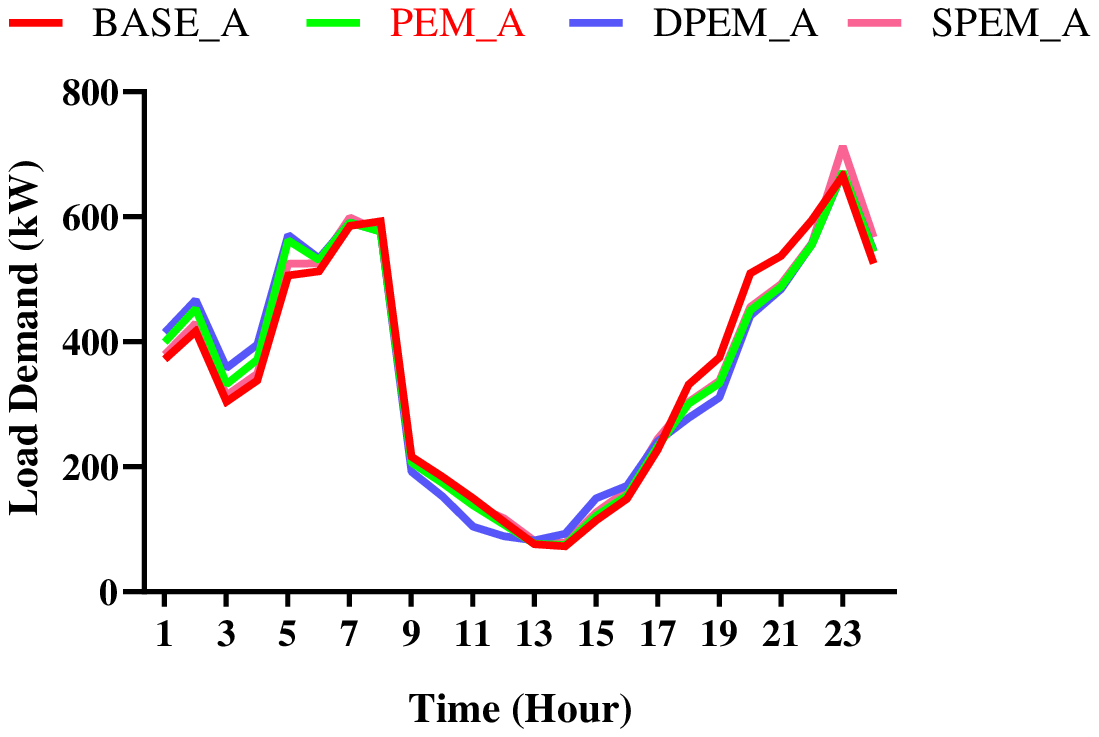}
     		\caption{}
     		\label{}
     	\end{subfigure}
     	\hfill
     	\caption{Aggregated class-wise demand profiles before DR as base case, and after DR using PEM, proposed DPEM and proposed SPEM: (a) Residential Customers, (b) large Industrial Customers, (c) Medium Industrial Customers, (d) Commercial Customers, (e) Agricultural Customers}
     	\label{aggloadall}
     \end{figure}
    Fig.~\ref{aggloadall} (a)-(e) shows the aggregated class-wise customer demand for base load BDR and ADR using the PEM, proposed DPEM and SPEM model. It can be observed from figure that PEM displays DR behaviour persuasively during peak hours. However, its recovery during off-peak/valley hours is not equal to the curtailed demand. It can be shown that R, C and MI customers exhibit partial increase in the demand during off-peak hours. Though, LI and A customers display much better adjustment of power during the off-peak hours. On the other hand, the proposed DPEM and SPEM display better curtailment and shifting for all the customer’s classes. In case of DPEM, dynamic elasticity makes curtailed load adaptive over the time-frame. It recovers the curtailed load in low price hours, indicating complete load recovery. This satisfies the one of assumption of the equivalent energy consumption BDR and ADR. In addition, the proposed DPEM is independent of customers’ load pattern showing its feasibility to all classes’ load pattern. Likewise, the proposed SPEM also exhibits better load recovery of the curtailed load using stochastic process accompanied by the intertemporal constraint. This may/may not maintain the energy balance constraint. But it displays feasible uncertainty associated with the customer response in DR.  \\
     \FloatBarrier
\begin{figure}[!h]
	\centering
	\begin{subfigure}[b]{0.45\textwidth}
		\centering
		\includegraphics[width=\textwidth]{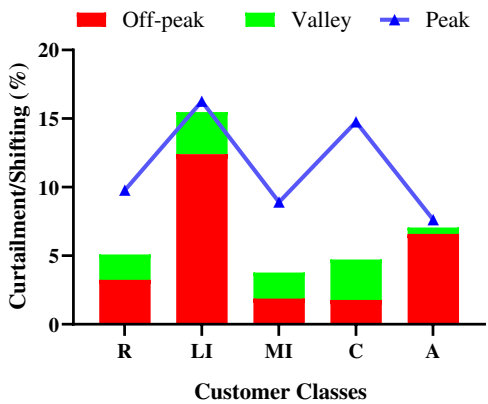}
		\caption{}
		\label{}
	\end{subfigure}
	\hfill
	\begin{subfigure}[b]{0.45\textwidth}
		\centering
		\includegraphics[width=\textwidth]{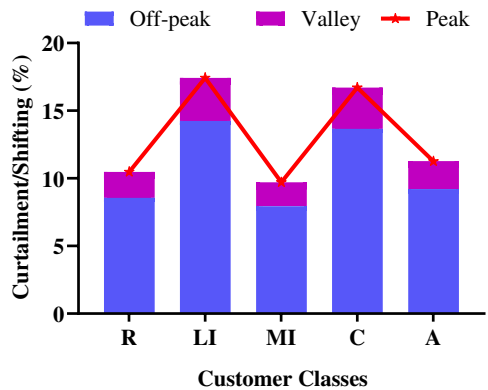}
		\caption{}
		\label{}
	\end{subfigure}
	\hfill
	\begin{subfigure}[b]{0.45\textwidth}
		\centering
		\includegraphics[width=\textwidth]{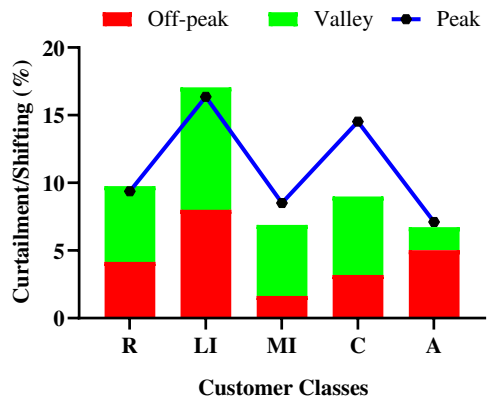}
		\caption{}
		\label{}
	\end{subfigure}
	\hfill
	\caption{Curtailed and shifted demand under RTP using (a) PEM, (b) DPEM and (c) SPEM.}
	\label{curshift}
\end{figure}

The aggregated DR in class-wise customers can be better illustrated from the bar graph as shown in Fig.~\ref{curshift}. These subfigures demonstrate the percentage change in the load demand BDR and ADR for the considered states. It can be observed from the Fig.~\ref{curshift} (a) that customers of R, C and MI class are partially able to shift the demand, while LI and A customers shift closely to the curtailed load demand. This partial shifting is due to the existence of high relative ratio between segmented periods as discussed earlier. Thus, the adaptiveness in the elasticity is important for load recovery, which can be fulfilled using the proposed DPEM. In DPEM, dynamic elasticity shows better adjustment of demand, where the percentage change in curtailed and shifted demand in each customers’ class is nearly equal as shown in Fig.~\ref{curshift} (b) as opposite to PEM. Similarly, SPEM gives better load recovery using stochastic process as can be observed from Fig.~\ref{curshift} (c). Though, it does not maintain equal energy consumption before and after DR due to uncertainty aspects. \\
 \begin{figure}[hbt!]
	\centering 
	\includegraphics[width=0.5\textwidth]{{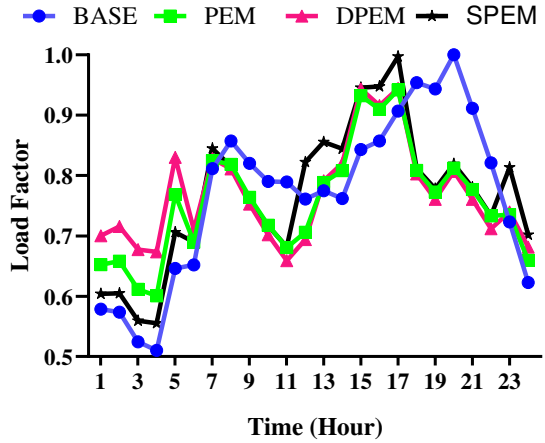}}
	\caption{Aggregated load factor at the utility level. }
	\label{loadfactor}
\end{figure}
The impact of DR on the system aggregated load factor is illustrated in Fig.~\ref{loadfactor}. The figure indicates that the load factor is considerably reduced during peak hours. On the other hand, significant improvement is found during off-peak/valley hours. The overall increase or decrease balanced out the difference of peak and off-peak/valley load factor. It gives an evenly distributed load factor displaying a uniform loading in the system. The figure reveals that the proposed PEM methods give load factor better or equivalent to standard PEM. From the economic point of view, an aggregated class-wise customer electricity bills BDR and ADR for the considered states using proposed DPEM framework are shown in Table.~\ref{tab:billstatewise}. \\
The table shows that peak hour bills of customers are increased even after participating in DR. Though, electricity bill during valley and off-peak period are decreased significantly. This indicates that customer need to curtail more load to come at break-even point during peak period. The aggregated class-wise bills are summarized in Table.~\ref{tab:classwisecost}. It can be observed from the table that R, C and MI customer class’ bills are increased after participating in DR, with highest increase in C and MI is next to it. This will discourage the customers to participate in DR. On the other hand, LI and A customer class’ electricity bills are decreased. Moreover, the overall increase in the electricity bills is around 0.50\%. \\
\FloatBarrier
\begin{table}[H]
\caption{Aggregated customer class bills under different price states using the proposed DPEM.}
\label{tab:billstatewise}
\begin{tabular}{|l|l|l|l|l|l|l|}
	\hline
	\multirow{2}{*}{Class} & \multicolumn{3}{c|}{BDR (¢)}      & \multicolumn{3}{c|}{ADR (¢)}      \\ \cline{2-7} 
	&
	\multicolumn{1}{c|}{Peak} &
	\multicolumn{1}{c|}{Valley} &
	\multicolumn{1}{c|}{Off-peak} &
	\multicolumn{1}{c|}{Peak} &
	\multicolumn{1}{c|}{Valley} &
	\multicolumn{1}{c|}{Off-peak} \\ \hline
	R                      & 115110.26 & 72379.75  & 56868.55  & 132888.01 & 67624.62  & 52412.90  \\ \hline
	LI                     & 210234.26 & 147044.53 & 195118.28 & 221252.16 & 140740.60 & 170027.95 \\ \hline
	MI                     & 32671.00  & 24710.61  & 11800.84  & 37757.22  & 23151.34  & 11615.31  \\ \hline
	C                      & 148834.80 & 109778.81 & 28230.88  & 163231.89 & 103621.78 & 36415.08  \\ \hline
	A                      & 36303.67  & 7825.74   & 43946.21  & 41393.43  & 7740.88   & 37137.38  \\ \hline
\end{tabular}
\end{table}      
\FloatBarrier
\begin{table}[H]
\caption{Aggregated class-wise customer electricity bills using the proposed DPEM.}
\label{tab:classwisecost}
\begin{tabular}{|l|l|l|l|l|l|l|}
	\hline
	Class &
	\multicolumn{1}{c|}{R} &
	\multicolumn{1}{c|}{LI} &
	\multicolumn{1}{c|}{MI} &
	\multicolumn{1}{c|}{C} &
	\multicolumn{1}{c|}{A} &
	\multicolumn{1}{c|}{Overall} \\ \hline
	BDR (¢)  & 244358.56 & 552397.07 & 69182.46 & 286844.49 & 88075.63 & 1240858.20 \\ \hline
	ADR (¢)  & 252925.53 & 532020.72 & 72523.87 & 303268.76 & 86271.69 & 1247010.56 \\ \hline
	Diff (¢) & 8566.97   & -20376.35 & 3341.41  & 16424.27  & -1803.94 & 6152.36    \\ \hline
	Change (\%) &
	\multicolumn{1}{c|}{3.51} &
	\multicolumn{1}{c|}{-3.69} &
	\multicolumn{1}{c|}{4.83} &
	\multicolumn{1}{c|}{5.73} &
	\multicolumn{1}{c|}{-2.05} &
	\multicolumn{1}{c|}{0.50} \\ \hline
\end{tabular}
\end{table}
\subsection{Comparative analysis}
\begin{figure}[hbt!]
	\centering 
	\includegraphics[width=0.42\textwidth]{{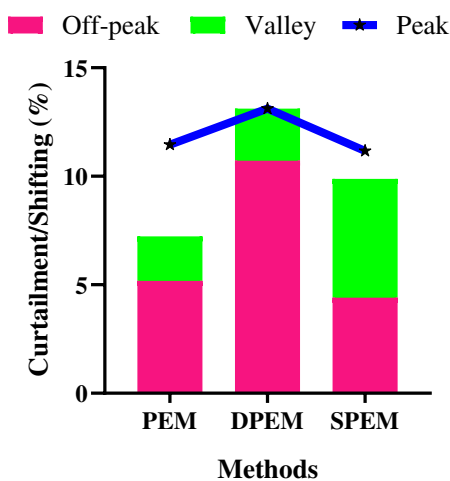}}
	\caption{Curtailed and shifted demand under RTP using PEM, the proposed DPEM and the proposed SPEM. }
	\label{curshiftmethods}
\end{figure}

{\label{com}}
This section investigates the usefulness of the proposed DPEM and SPEM with the existing PEM on an aggregated scale. Therefore, the overall response in DR using proposed and existing methods is shown in Fig.~\ref{curshiftmethods}. This figure shows that all the methods demonstrate peak curtailment, effectively. However, the absorption of this curtailed demand exhibits varied response over the off-peak/valley hours under the different methods. Where PEM exhibits moderate absorption of the curtailed load, whereas DPEM displays the complete absorption using adaptive elasticity. However, both PEM and proposed DPEM adjust its most of curtailed demand to the off-peak period as shown in figure, which contradicts with the theory of distinct elasticity pattern for the relative increment/decrement in the price. This assumption is further accompanied by the customer’s intertemporal constraint of load flexibility, which diminishes non-linearly over the cross-periods (i.e., load recovery decreases as subsequent cross-periods are further away from peak hours). Thus, these two assumptions make PEM and proposed DPEM lack in incorporating intertemporal flexibility constraint for the load recovery. On the contrary, the proposed SPEM incorporates the flexibility constraint as stochastic process using GBM. This gives realistic load recovery with the element of uncertainty. Further, it provides an evenly shifted load in the valley and off-peak periods. \\
 \\
In addition, a comparative cost assessment is also performed for PEM, proposed DPEM and SPEM as compiled in Table.~\ref{tab:comcost}. The results indicate that the total customers’ bill decreases in case of standard PEM, whilst increases for DPEM and SPEM. Though, decrease in the bills for PEM is at the expense of the lowered energy consumption ADR, which may cause discomforts to the customers. The bill increase in SPEM is slightly higher than DPEM. It is due to pragmatic load shifting in the valley/off-peak periods with the intertemporal bounds. It is worth to mention that the presented economic analysis gives a snapshot for the considered RTP signal. Thus, the results may vary, when different RTP signal is employed. \\
\FloatBarrier
\begin{table}[H]
\caption{Comparative cost analysis under the different DR models.}
\label{tab:comcost}
\begin{tabular}{|l|l|l|c|c|}
	\hline
	Methods & Total   Cost (¢) & Diff   (¢) & \multicolumn{1}{l|}{Change (\%)} & \multicolumn{1}{l|}{Total Energy (kWh) (\%)} \\ \hline
	Base            & 1240858.20 & \multicolumn{1}{c|}{-} & -     & 59402.00 (-)     \\ \hline
	PEM             & 1235350.91 & -5507.29               & -0.44 & 58553.95 (-1.43) \\ \hline
	Proposed   DPEM & 1247010.56 & 6152.36                & 0.50  & 59402.00 (0)     \\ \hline
	Proposed   SPEM & 1257491.27 & 22140.36               & 1.34  & 59265.61 (-0.23) \\ \hline
\end{tabular}
\end{table}
\section{Conculsions}
This paper presents an adaptive economical DR framework using PEM to model DR in the distribution networks. The proposed model emulates the key features of DR (adaptability and adjustability) through a dynamic elasticity using deterministic and stochastic approaches, which was partially present in the existing PEM. Both approaches incorporate dynamism in the elasticity to make load recovery in the cross-periods. In DPEM, the proposed dynamic elasticity inherently exhibits the traits of both self and cross-elasticity, and establishes an interlink between peak, valley and off-peak period for load shifting. It demonstrates that whatever the amount of demand is curtailed during peak hours will be shifted to off-peak hours irrespective of customers’ load pattern. Though, the proposed DPEM exhibits true DR, but it lacks in incorporating intertemporal constraint of load flexibility and the virtue of diminishing load recovery. This is effectively overcome in the proposed SPEM, which imitate a feasible load recovery using GBM to assess the uncertain behaviour of customers in DR. It can be corroborated from the result analysis that DPEM is more optimistic model, which may provide the overestimated results in the practical situation. On the other hand, SPEM is a progressive framework due to its likeness to realistic condition. Though, its effectiveness depends upon the quantification of uncertainty of customer’s non-linear behaviour in DR. The one of the outcomes of the economic analysis illustrates that participating in DR programs will not be beneficial for all customers due to the customers’ heterogeneous behaviour. Since, the PEM is governed by the price and demand pattern. Thus, the present work can be extended in future to observe customer participation level in DR, when the price pattern is designed according to the customer class load pattern. \\
\bibliographystyle{unsrt}

\bibliography{cas-refs}

\end{document}